\documentclass[ALICE,manyauthors]{cernphprep}
\usepackage{hyperref}

\newcommand{\pt}{$p_{\mathrm t}$}
\usepackage{rotating}
\begin{document}%
%
%
\begin{titlepage}
\PHnumber{2011-215}      
\PHdate{22 December 2011}              
%
%
\title{Heavy flavour decay muon production at forward rapidity 
in proton--proton collisions at $\sqrt s$~=~7~TeV} 
\ShortTitle{Heavy flavour decay muon production in pp collisions at 
$\sqrt s$ = 7 TeV}
%
\Collaboration{ALICE Collaboration%
         \thanks{See Appendix~\ref{app:collab} for the list of collaboration 
                      members}}
\ShortAuthor{ALICE Collaboration}  
%
\begin{abstract}
The production of muons from heavy flavour decays is measured 
at forward rapidity in proton--proton collisions at $\sqrt s$~=~7~TeV 
collected with the ALICE experiment at the LHC. The analysis is carried out 
on a data sample corresponding to an integrated luminosity 
$L_{\rm int}$~=~16.5~nb$^{-1}$. The transverse momentum and rapidity 
differential production cross sections of muons from heavy flavour decays are 
measured in the rapidity range $2.5 < y < 4$, over the transverse momentum 
range $2 < p_{\mathrm t} < 12$~GeV/$c$. The results are compared to 
predictions based on perturbative QCD calculations. 

\vspace*{1 cm}
\end{abstract}

\end{titlepage}
\setcounter{page}{2}
\section{Introduction}

The study of heavy flavour (charm and beauty) production in proton--proton 
collisions at LHC (Large Hadron Collider) energies provides an important test 
of perturbative QCD (pQCD) calculations~\cite{mnr92,FONLL1} in a new 
energy domain, where unprecedented small Bjorken-$x$ (momentum fraction) 
values are probed. In the rapidity region $2.5 < y < 4$, charm (beauty) 
production at $\sqrt s$ = 7 TeV is expected to be sensitive to $x$ values 
down to about $6 \cdot 10^{-6}$ ($2 \cdot 10^{-5}$). Important progress 
has been achieved in the understanding of heavy flavour production at lower 
energies. In earlier measurements, the beauty production cross section 
in p$\rm \overline{p}$ collisions at $\sqrt s$ = 1.8 TeV measured by 
the CDF and D0 experiments~\cite{d01,cdf1} at the FNAL Tevatron, 
was found to be higher than Next to Leading Order (NLO) 
pQCD predictions~\cite{mnr92}. More recent results from the CDF 
Collaboration~\cite{cdf2}, for p$\rm \overline{p}$ collisions at 
$\sqrt s$ = 1.96 TeV, are described well by Fixed Order Next-to-Leading Log 
(FONLL)~\cite{cac02,cac04} and NLO~\cite{kni08} pQCD calculations. The 
charm production cross section measured at the FNAL Tevatron~\cite{aco03} is 
also well reproduced by FONLL~\cite{cac03} and GM-VFN~\cite{kni06} 
calculations within experimental and theoretical uncertainties, although 
at the upper limit of the calculations. 
The PHENIX and STAR Collaborations~\cite{phe1,sta1} at the 
RHIC (Relativistic Heavy Ion Collider) measured the production of muons and 
electrons from heavy flavour decays in pp collisions at $\sqrt s$~=~0.2 TeV. 
The upper limit of FONLL pQCD calculations~\cite{cac05} is consistent with 
the measurement of electrons from heavy flavour decays in the mid-rapidity 
region, while in the forward rapidity region the production of muons from 
heavy flavour decays is underestimated by the model calculations. 
Furthermore, at LHC energies, the ATLAS~\cite{atlas1}, LHCb~\cite{lhcb1} 
and CMS~\cite{cms1,cms2} Collaborations reported on the measurement of 
beauty production in pp collisions at $\sqrt s$~=~7~TeV. The results 
are consistent with NLO pQCD calculations within uncertainties. 
A similar agreement with FONLL calculations is also observed for mid-rapidity 
electrons and muons from heavy flavour decays, measured by the 
ATLAS experiment~\cite{atlas2} in pp collisions at $\sqrt s$~=~7~TeV. 
In this respect, it is particularly interesting to perform the measurement 
of heavy flavour decay muon production in the forward rapidity region 
at the LHC and compare it with theoretical models.

The investigation of heavy flavour production in pp collisions 
also constitutes an essential baseline for the corresponding measurements 
in heavy-ion collisions. In the latter, heavy quarks are produced at early 
stages of the collision and then experience the full evolution of the 
extremely hot and dense, strongly interacting medium~\cite{car04,ale06}. 
The modification of the heavy flavour 
transverse momentum distributions measured in heavy ion collisions 
with respect to those measured in pp collisions is considered as a sensitive 
probe of this medium~\cite{arm05,djo05}. 

Finally, the study of heavy flavour production is also important for the 
understanding of quarkonium production, both in pp, p--nucleus and 
nucleus--nucleus collisions~\cite{car04,ale06}.

The ALICE experiment~\cite{aam08} measures the heavy flavour production
at mid-rapidity through the semi-electronic decay channel~\cite{HFmid11} 
and in a more direct way through the hadronic D-meson decay 
channel~\cite{Dmes11}, and at forward rapidity through the semi-muonic 
decay channel. In this paper, we present the measurement of 
differential production cross sections of muons from heavy flavour 
decays in the rapidity range $2.5 < y < 4$ and transverse momentum range 
$2 <p_{\rm t} < 12$ GeV/$c$, with the ALICE muon spectrometer~\cite{aam08}, 
in pp collisions at $\sqrt s$~=~7~TeV. The results are compared 
to FONLL pQCD calculations~\cite{FONLL1, FONLL3}. 

The paper is organized as follows. Section 2 consists of an overview 
of the ALICE experiment with an emphasis on the muon spectrometer 
and a description of data taking conditions. Section 3 is devoted to the 
analysis strategy: event and track selection, background subtraction, 
corrections, normalization and determination of systematic uncertainties. 
Section 4 addresses the experimental results: \pt- and $y$-differential 
production cross sections of muons from heavy flavour 
decays at forward rapidity, and comparisons to FONLL pQCD predictions. 
Conclusions are given in Section~5.

\section{The ALICE experiment and data taking conditions}

A detailed description of the ALICE detector can be found in~\cite{aam08}. 
The apparatus consists of two main parts: a central barrel (pseudo-rapidity 
coverage: $\vert \eta \vert < 0.9$) placed in a large solenoidal 
magnet ($B$~=~0.5~T), which measures hadrons, electrons and photons, 
and a muon spectrometer ($-4 < \eta < -2.5$\footnote{The muon spectrometer 
covers a negative pseudo-rapidity range in the ALICE reference frame. 
$\eta$ and $y$ variables are identical for muons in the acceptance of the muon 
spectrometer, and in pp collisions the physics results are symmetric 
with respect to $\eta$ ($y$) = 0. They will be presented as a function 
of $y$, with positive values.}). Several smaller detectors for global event 
characterization and triggering are located in the forward and backward 
pseudo-rapidity regions. Amongst those, the VZERO detector is used 
for triggering purposes and in the offline rejection of beam-induced 
background events. It is composed of two scintillator arrays placed at 
each side of the interaction point and covering 
$2.8 < \eta < 5.1$ and $-3.7 < \eta < -1.7$. The central barrel detector 
used in this work for the interaction vertex measurement is the 
Silicon Pixel Detector (SPD), the innermost part of the 
Inner Tracking System (ITS). The SPD consists of two cylindrical 
layers of silicon pixels covering $\vert \eta \vert < 2.0$ and 
$\vert \eta \vert < 1.4$ for the inner and outer layer, respectively. 
The SPD is also used in the trigger logic. 

The muon spectrometer detects muons with momentum larger than 4 GeV/$c$ 
and is composed of two absorbers, a dipole magnet providing a field integral 
of 3~Tm, and tracking and trigger chambers. A passive front absorber 
of 10 interaction lengths ($\lambda_{\rm I}$), made of carbon, 
concrete and steel, is designed to reduce the contribution of hadrons, 
photons, electrons and muons from light hadron decays. A small angle beam 
shield ($\theta < 2^{\rm o}$), made of tungsten, lead and steel, protects 
the muon spectrometer against secondary particles produced by the interaction 
of large-$\eta$ primary particles in the beam pipe. Tracking is performed 
by means of five tracking stations, each composed of two planes of 
Cathode Pad Chambers. Stations 1 and 2 (4 and 5) are located upstream 
(downstream) of the dipole magnet, while station 3 is embedded inside 
the dipole magnet. The intrinsic spatial resolution of the tracking chambers 
is better than 100~$\mu$m. Two stations of trigger chambers equipped with 
two planes of Resistive Plate Chambers each are located downstream of the 
tracking system, behind a 1.2~m thick iron wall of 7.2 $\lambda_{\rm I}$. 
The latter absorbs most of the hadrons that punch through the front absorber, 
secondary hadrons produced inside the front absorber and escaping it 
and low momentum muons ($p < 4$ GeV/$c$). The spatial resolution of the 
trigger chambers is better than 1~cm and the time resolution is about 2~ns. 
Details concerning track reconstruction can be found in~\cite{jpsi,aph09}.

The results presented in this publication are based on the analysis of a 
sample of pp collisions at $\sqrt s$~=~7~TeV collected in 2010, corresponding 
to an integrated luminosity of 16.5 nb$^{-1}$.

The data sample consists of minimum bias trigger events (MB) and muon trigger 
events ($\mu$-MB), the latter requiring, in addition to the MB trigger 
conditions, the presence of one muon above a transverse momentum (\pt) 
threshold that reaches the muon trigger system. The MB trigger is 
defined as a logical OR between the requirement of at least one hit in the 
SPD and a hit in one of the two VZERO scintillator arrays. 
It also asks for a coincidence between the signals from the two beam counters, 
one on each side of the interaction point, indicating the passage of bunches. 
This corresponds to at least one charged particle in 8 units of 
pseudo-rapidity. The logic of the $\mu$-MB trigger requires hits in at least 
three (out of four possible) trigger chamber planes. The estimate of the muon 
transverse momentum is based on the deviation of the measured track with 
respect to a straight line coming from the interaction point, in the bending 
plane (plane measuring the position along the direction perpendicular to the 
magnetic field). By applying a cut on this deviation, tracks above 
a given \pt\ threshold are selected. The \pt\ threshold allows the 
rejection of soft background muons mainly coming from pion and kaon decays, 
and also to limit the muon trigger rate when high luminosities are delivered 
at the interaction point. In the considered data taking period, the \pt\ 
trigger threshold was set to its minimum value of about 0.5 GeV/$c$ 
and the corresponding muon trigger rate varied between about 40 and 150 Hz. 
The instantaneous luminosity at the ALICE interaction point was limited to 
0.6--1.2$\cdot$10$^{29}$ cm$^{-2}$ s$^{-1}$ by displacing the 
beams in the transverse plane by 3.8 times the r.m.s of their transverse 
profile. In this way, the probability to have multiple MB interactions 
in the same bunch crossing is kept below 2.5$\%$. 

The alignment of the tracking chambers, a crucial step for the single muon 
analysis, was carried out using the MILLEPEDE package~\cite{blo02}, by 
analyzing tracks without magnetic field in the dipole and solenoidal 
magnet. The corresponding resolution is about 300 $\mu$m in the 
bending plane, for tracks with \pt\ ~$>$~2~GeV/$c$. With such alignment 
precision, the relative momentum resolution of reconstructed tracks ranges 
between about 1$\%$ at a momentum of 20 GeV/$c$ and 4$\%$ at 100 GeV/$c$.

\section{Data analysis}

\begin{figure}[ht!]
\centering
{\includegraphics[width=.6\textwidth]{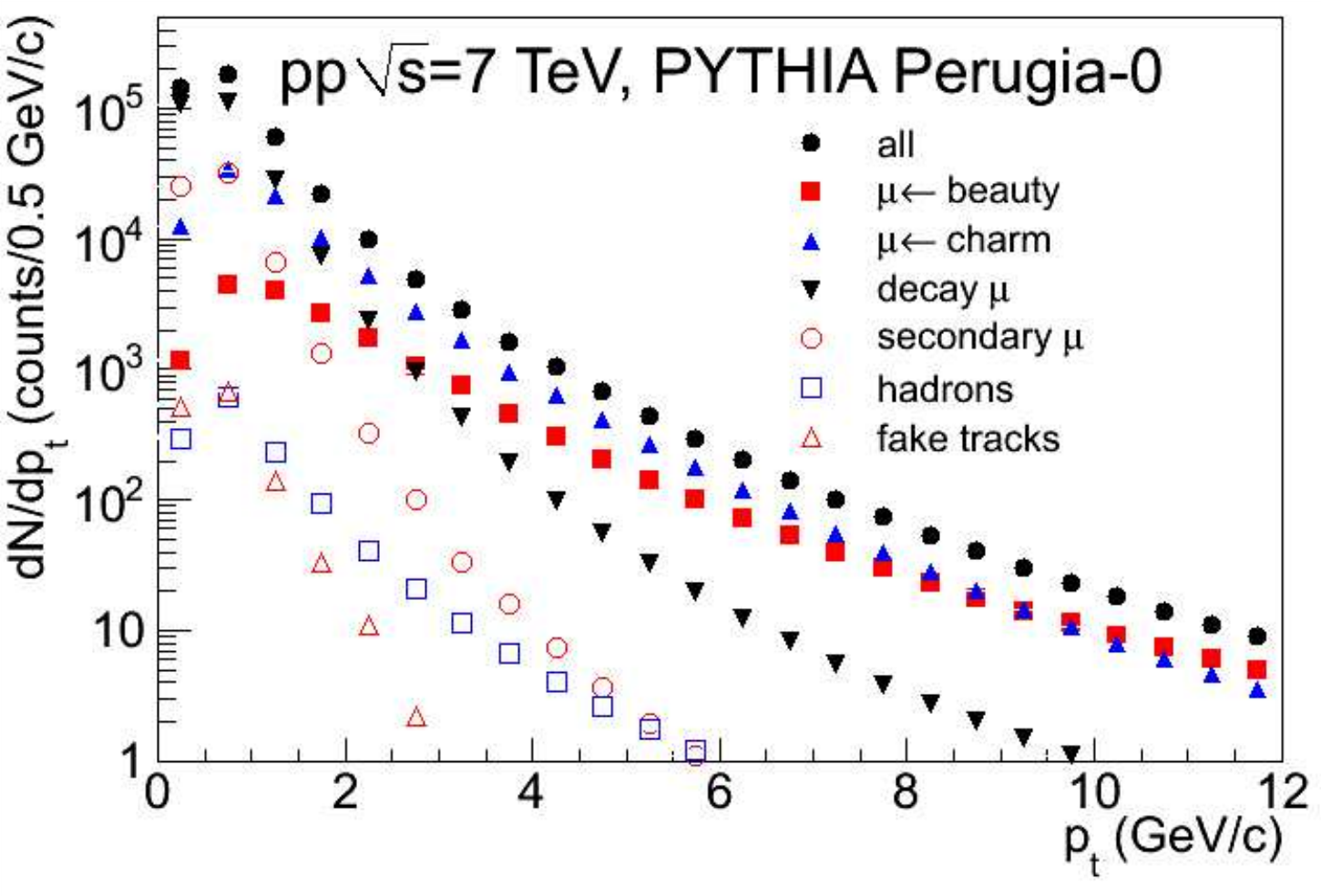}}
\caption{Transverse momentum distribution of reconstructed 
tracks in the muon spectrometer after all selection cuts were applied 
(see section 3.1 for details). The distributions were obtained from a 
PYTHIA~\cite{pyt1,pyt2} (tune Perugia-0~\cite{per10}) simulation of pp 
collisions at $\sqrt s$~=~7 TeV. 
The main sources are indicated in 
the figure.}
\label{fig:sim}
\end{figure}

The single muon analysis was carried out with muon trigger events while, as 
will be discussed in Section~3.4, minimum bias trigger events were used to 
convert differential muon yields into differential cross sections. 
The identification of muons from charm and beauty decays in the 
forward region is based on the \pt\ distribution of reconstructed tracks. 
Three main background contributions must be subtracted and/or rejected: 
\begin{itemize}
\item decay muons: muons from the decay of primary light hadrons 
including pions and kaons 
(the main contribution) and other meson and baryon decays 
(such as $\rm J/\psi$ and low mass resonances $\eta$, $\rho$, $\omega$ 
and $\phi$);
\item secondary muons: muons from secondary light hadron decays produced 
inside the front absorber;
\item punch-through hadrons and secondary hadrons escaping the 
front absorber and crossing the tracking chambers, which are wrongly 
reconstructed as muons.
\end{itemize}
A Monte Carlo simulation based on the GEANT3 transport 
code~\cite{geant3,geant3b} and 
using the PYTHIA 6.4.21 event generator~\cite{pyt1,pyt2} 
(tune Perugia-0~\cite{per10}) was performed to 
obtain the \pt\ distributions of these different contributions. They are 
displayed in Fig.~\ref{fig:sim} after all the selection cuts discussed 
in Section 3.1 were applied. After cuts, the component of muons 
from heavy flavour decays prevails over the background contribution for 
\pt\ $\gtrsim$ 4 GeV/$c$. The simulation results indicate that the hadronic 
background and the contribution of fake tracks (tracks which are not 
associated to one single particle crossing the whole spectrometer) are 
negligible. The component of muons from W$^\pm$ and Z$^0$ decays, 
which dominates in the \pt\ range 30--40 GeV/$c$~\cite{zai07,atlas2}, is 
not considered in this analysis. This contribution is negligible in 
the \pt\ range of interest 2--12~GeV/$c$. 

\subsection{Data sample: event and track selection}

The data sample used in the physics analysis amounts to 1.3$\cdot$10$^7$ 
$\mu$-MB trigger events. These selected events satisfied the quality criteria 
on detector conditions during data taking and the analysis quality criteria, 
which reduced the beam-induced background. This was achieved 
by using the timing information from the VZERO and by exploiting the 
correlation between the number of hits and track segments 
in the SPD. The accepted events have at least one interaction vertex 
reconstructed from hits correlation in the two SPD layers. The corresponding 
total number of tracks reconstructed in the muon spectrometer is 
7.8$\cdot$10$^6$. Various selection cuts were applied in order to reduce 
the background contributions in the data sample. Tracks were required to be 
reconstructed in the geometrical acceptance of the muon spectrometer, 
with $-4 < \eta < -2.5$ and $171^{\rm o} < \theta_{\rm abs} < 178^{\rm o}$, 
$\theta_{\rm abs}$ being the track polar angle measured at the end of the 
absorber. These two cuts reject about 9$\%$ of tracks. Then, the track 
candidate measured in the muon tracking chambers was required to be matched 
with the corresponding one measured in the trigger chambers. 
This results in a very effective rejection of the hadronic component that is 
absorbed in the iron wall. This condition is fulfilled for a large fraction 
of reconstructed tracks since the analysis concerns $\mu$-MB trigger events. 
The fraction of reconstructed tracks that are not matched with a corresponding 
one in the trigger system is about 5$\%$. For comparison, in MB collisions 
this fraction is about 64$\%$. Furthermore, the correlation between momentum 
and Distance of Closest Approach (DCA, distance between the extrapolated 
muon track and the interaction vertex, in the plane perpendicular to the 
beam direction and containing the vertex) was used to remove remaining 
beam-induced background tracks which do not point to the interaction vertex. 
Indeed, due to the multiple scattering in the front absorber, 
the DCA distribution of tracks coming from the interaction vertex is 
expected to be described by a Gaussian function whose width depends on the 
absorber material and is proportional to $1/p$. The beam-induced background 
does not follow this trend and can be rejected by applying a cut on 
$p \times$~DCA at 5$\sigma$, where $\sigma$ is extracted from a Gaussian 
fit to the $p \times$~DCA distribution measured in two regions 
in $\theta_{\rm abs}$, corresponding to different materials in the front 
absorber. This cut removes 0.4$\%$ of tracks, mainly located in the 
high \pt\ range (in the region \pt\ $>$ 4~GeV/$c$, this condition rejects 
about 13$\%$ of tracks). After these cuts, the data sample consists 
of 6.67$\cdot$10$^6$ muon candidates. 

The measurement of the heavy flavour decay muon production is performed 
in the region \pt\ $>$~2 GeV/$c$ where the contribution of secondary muons 
is expected to be small (about 3$\%$ of the total muon yield, see 
Fig.~\ref{fig:sim}). In such a \pt\ region the main background 
component consists of decay muons and amounts to 
about 25$\%$ of the total yield (see Fig.~\ref{fig:sim}).

\subsection{Subtraction of the background contribution of decay muons}

The subtraction of the background component from decay muons (muons from 
primary pion and kaon decays, mainly) is based on simulations, using  
PYTHIA 6.4.21~\cite{pyt1,pyt2} (tune Perugia-0~\cite{per10}) 
and PHOJET 1.12~\cite{pho} as event generators. In order to avoid fluctuations 
due to the lack of statistics in the high \pt\ region in the Monte Carlo 
generators, the reconstructed \pt\ distribution of decay muons, 
obtained after all selection cuts are applied (Section 3.1), is fitted using 
\begin{equation}
{{\rm d}N\over {{\rm d} p_{\rm t}}}^{\mu\leftarrow {\rm decay}} = 
{{\rm a} \over {(p_{\rm t}^2 +  {\rm b})^{\rm c}}},
\end{equation}
where a, b and c are free parameters. 
The fits are performed in five rapidity intervals, 
in the region $ 2.5 < y < 4$. The normalization is done assuming that the 
fraction of decay muons in the data is the same as the one in the simulations, 
in the region where this component is dominant ($p_{\rm t}$~$<$~1 GeV/$c$). 
Finally, the (fitted) \pt\ distribution is subtracted from the measured 
muon \pt\ distribution. The subtracted \pt\ distribution is the mean of 
the \pt\ distributions from the PYTHIA and PHOJET event generators. 

The total systematic uncertainty due to this procedure 
includes contributions from the model input and the transport code 
(GEANT3~\cite{geant3,geant3b}). The former takes into account the shape 
and normalization of the \pt\ distribution of decay muons, and 
the observed difference in the $\rm K^\pm/\pi^\pm$ ratio as a function 
of \pt\ in the mid-rapidity region~\cite{cho11} between ALICE data and 
simulations. The results show that both PYTHIA (tune Perugia-0) and 
PHOJET underestimate this ratio by about 20$\%$. The corresponding 
uncertainty due to this difference between data and simulations is 
propagated to the muon yield in the forward rapidity region. 
The effect of the transport code is 
estimated by varying the yield of secondary muons within 100$\%$ in 
such a way to provide a conservative estimate of the systematic 
uncertainty on the secondary particle production in the front absorber. 
The systematic uncertainty from the model input varies from about 7$\%$ to 
2$\%$ as $y$ increases from 2.5 to 4, independently of \pt, while the one 
from the transport code depends both on $y$ and \pt\ 
and ranges from 4$\%$ ($3.7 < y< 4$) to a maximum of 34$\%$ 
(\pt\ = 2 GeV/$c$ and $2.5 < y < 2.8$). The corresponding 
values of these systematic uncertainties as a function of \pt\ and $y$ are 
summarized in Table~\ref{tab-systbck1}. They are added in quadrature in 
the following.

\begin{table}
\caption{\label{tab-systbck1} Systematic uncertainties 
introduced by the procedure used for the subtraction of decay muons. 
MC and transport refer to 
the systematic uncertainty due to model input and transport code, 
respectively. See the text for details. }
\centering
\begin{tabular}{c|c|c|c|c|c|c|c|c}
\hline
& MC  & \multicolumn{7}{c} {transport} \\
\cline{3-9}
  & & \multicolumn{7}{c} {$p_{\rm t}$ (GeV/$c$)} \\
\cline{3-9} 
& & [2.0;
2.5] & [2.5;3.0] & [3.0;3.5] & [3.5;4.0] & [4.0;4.5] & 
[4.5;5.0] & 
$>$ 5.0 \\ \hline
$2.5 < y < 2.8$ & 7$\%$ & 34$\%$ & 22$\%$ & 20$\%$ & 16$\%$ & 12$\%$ 
& 10$\%$ & 6$\%$ \\ \hline
$2.8 < y < 3.1$ &  5.5$\%$ & 22$\%$ & 18$\%$ & 14$\%$ & 12$\%$ & 10$\%$ & 
8$\%$ & 6$\%$ \\ \hline
$3.1 < y < 3.4$ & 4.5$\%$ & 10$\%$ & 9$\%$ & 8$\%$ & 7$\%$ & 
\multicolumn{3}{c} {6$\%$} \\ \hline
$3.4 < y < 3.7$ & 3.0$\%$ &  \multicolumn{5}{c} {6$\%$} \\ \hline
$3.7 < y < 4.0$ & 2.0$\%$ &  \multicolumn{5}{c} {4$\%$} \\ \hline
\end{tabular}
\end{table}

\subsection{Corrections}

\begin{figure}[htbp]
\centering
{\includegraphics[width=.45\textwidth]{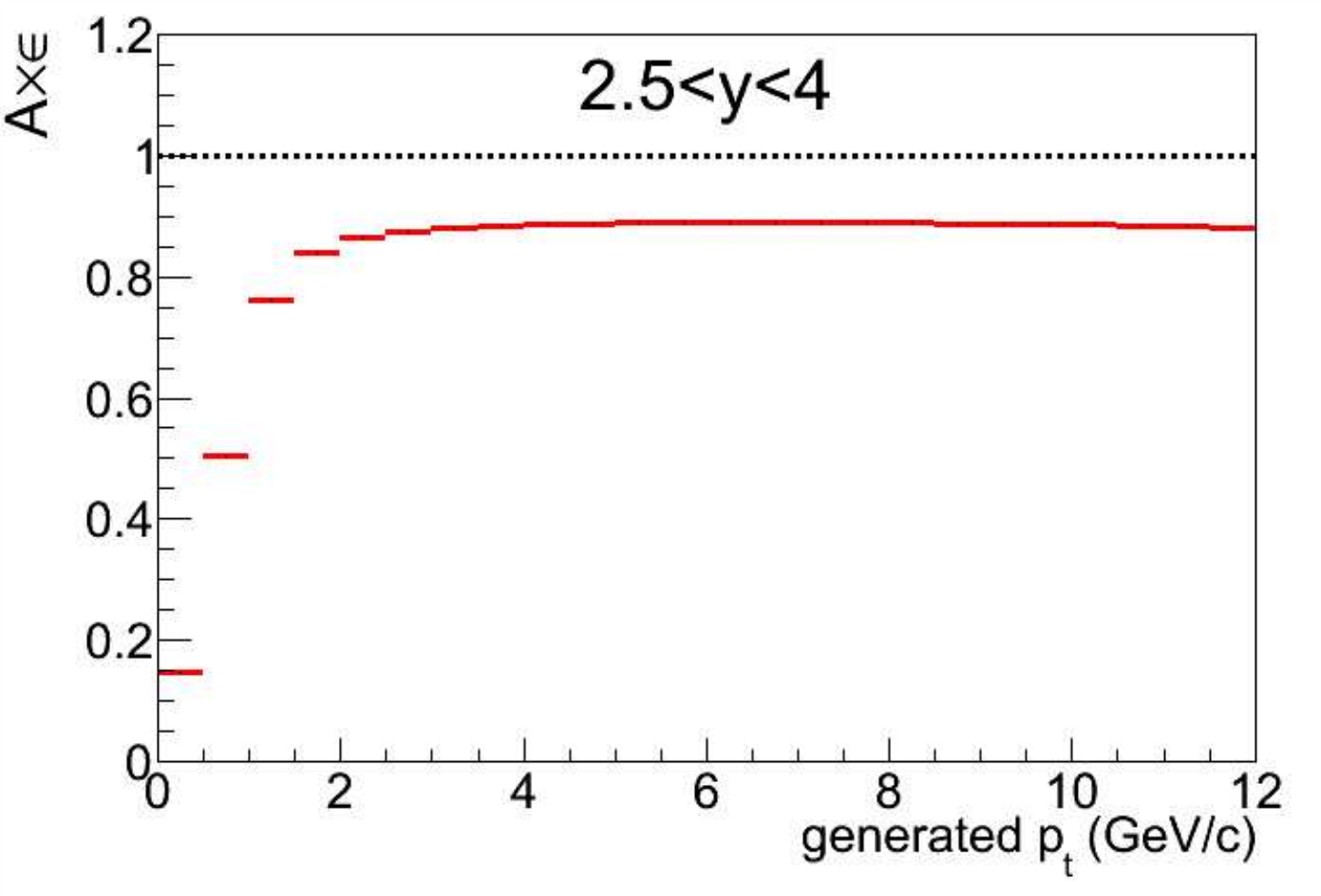}}
\caption{Acceptance $\times$ efficiency 
as a function of generated \pt, obtained from a simulation of muons 
from beauty decays.}
\label{fig:eff}
\end{figure}

The extracted yields of muons from heavy flavour decays are corrected for 
acceptance, reconstruction and trigger efficiencies by means of a 
simulation modelling the response of the muon spectrometer. The procedure 
is based on the generation of a large sample of muons from beauty decays by 
using a parameterization of NLO pQCD calculations~\cite{aph09}. 
The tracking efficiency takes into account the status of each 
electronic channel and the residual mis-alignment of detection elements. 
The evolution of the tracking efficiency over time is controlled by weighting 
the response of electronic channels as a function of time. The typical value 
of muon tracking efficiency is about 93$\%$. The efficiencies of the 
muon trigger chambers are obtained directly from data~\cite{jpsi} 
and employed in the simulations. The typical value of such efficiencies is 
about 96$\%$. Figure~\ref{fig:eff} shows the resulting acceptance and 
efficiency ($A \times \epsilon$) as a function of generated \pt.
The global $A \times \epsilon$ increases significantly up to about 
2 GeV/$c$ and tends to saturate at a value close to 90$\%$.

The systematic uncertainty corresponding to the sensitivity of 
$A \times \epsilon$ on the input \pt\ and $y$ distributions was estimated 
by comparing the results with those from a simulation using muons from charm 
decays. This amounted to less than 1$\%$ and was neglected. 
The accuracy in the detector modelling introduces a systematic uncertainty 
estimated to be 5$\%$, by comparing the values of trigger and tracking 
efficiencies extracted from data and simulations~\cite{jpsi}. 

The distortion of the measured \pt\ distribution, dominated in the 
high \pt\ region by the effect of residual mis-alignment, is also 
corrected for by introducing in the simulation a residual mis-alignment of 
the same order of magnitude as in the data. However, this residual 
mis-alignment is generated randomly. A \pt\ dependent relative systematic 
uncertainty on the muon yield of 1$\%$ $\times$ \pt\ (in GeV/$c$) is 
considered in order to take into account the differences between 
the real (unknown) residual mis-alignment and the simulated one. 
This is a conservative value determined by comparing the reconstructed 
\pt\ distribution with or without including the residual mis-alignment.

\subsection{Production cross section normalization}

The differential production cross section is obtained by normalizing 
the corrected yields of muons from heavy flavour decays to the integrated 
luminosity. Since the yields have been extracted using $\mu$-MB trigger 
events, the differential production cross section is calculated according to:
\begin{equation}
{{\rm d}^2 \sigma_{\rm \mu^\pm \leftarrow HF} 
\over{{\rm d} p_{\rm t}{\rm d} y} }= 
 {{\rm d}^2 N_{\mu^\pm\leftarrow \rm{HF}}
\over{{\rm d} p_{\rm t} {\rm d} y}} \times 
{N^{\mu^\pm}_{\rm MB} \over { N^{\mu^\pm}_{\mu-\rm {MB}}}} 
\times {\sigma_{\rm MB} \over {N_{\rm MB}}},
\end{equation}
where:
\begin{itemize}
\item ${{\rm d}^2 N_{\mu^\pm\leftarrow {\rm HF}}
\over{{\rm d} p_{\rm t} {\rm d} y }}$ is the \pt\ and $y$ differential 
yield of muons from heavy flavour decays;
\item $N^{\mu^\pm}_{\rm MB}$ 
and $N^{\mu^\pm}_{\mu-\rm {MB}}$ are the numbers of reconstructed 
tracks that satisfy the analysis cuts in MB and $\mu$-MB trigger events, 
respectively;
\item $N_{\rm MB}$ is the number of minimum bias collisions corrected 
as a function of time by the probability to have multiple MB interactions 
in a single bunch crossing, and 
$\sigma_{\rm MB}$ is the corresponding measured minimum bias cross section.
\end{itemize}

$\sigma_{\rm MB}$ is derived from the $\sigma_{\rm VZERO-AND}$ 
cross section~\cite{gag11} measured with the van der Meer scan 
method~\cite{vdm}. The $\rm VZERO-AND$ condition is defined as a logical 
AND between signals in the two VZERO scintillator arrays. 
Such a combination allows one to reduce the sensitivity to beam-induced 
background. The $\sigma_{\rm VZERO-AND}$/$\sigma_{\rm MB}$ ratio is the 
fraction of minimum bias events where the $\rm VZERO-AND$ condition is 
fulfilled. Its value is 0.87 and it remains stable within 1$\%$ over the 
analyzed data sample. This gives 
$\sigma_{\rm MB}$ = 62.5 $\pm$ 2.2 (syst.)~mb. The statistical 
uncertainty is negligible, while the 3.5$\%$ systematic uncertainty is 
mainly due to the uncertainty on the beam intensities~\cite{beam} and on 
the analysis procedure related to the van der Meer scan of 
the $\rm VZERO-AND$ signal. Other effects, such as oscillation in 
the ratio between MB and $\rm VZERO-AND$ counts, contribute less than 1$\%$. 

\subsection{Summary of systematic uncertainties}

The systematic uncertainty on the measurements of the \pt\ and $y$ 
differential production 
cross sections of muons from heavy flavour decays accounts 
for the following contributions discussed in the previous sections:
\begin{itemize}
\item background subtraction: from about 5$\%$ ($3.7 < y < 4$) 
to a maximum of 35$\%$ ($ 2.5 < y < 2.8$, \pt~=~2~GeV/$c$), 
see Section 3.2 and Table~\ref{tab-systbck1};
\item detector response: 5$\%$ (Section 3.3);
\item residual mis-alignment: 1$\%$ $\times$ \pt\ (Section 3.3); 
\item luminosity measurement: 3.5$\%$ (Section 3.4).
\end{itemize}

The resulting systematic uncertainty, in the rapidity 
region $ 2.5 < y < 4 $, varies between 8-14$\%$ (the 3.5$\%$ systematic 
uncertainty on the normalization is not included).

\section{Results and model comparisons}

\begin{figure}[htbp]
\centering
{\includegraphics[width=1.0\textwidth]{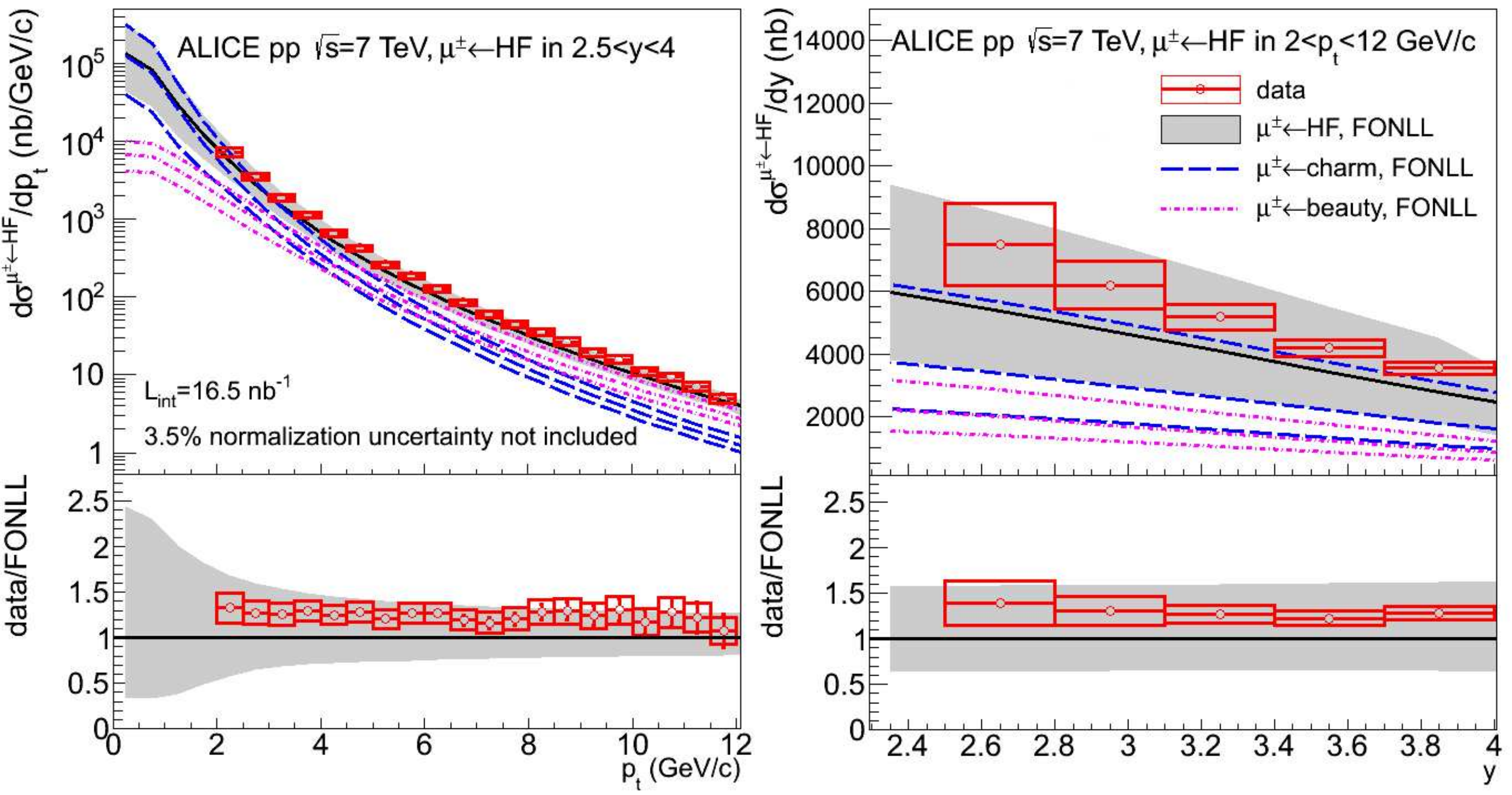}}
\caption{Left: \pt -differential production cross section of muons from 
heavy flavour decays in the rapidity range $2.5 < y < 4$. Right: 
y-differential production cross section of muons from 
heavy flavour decays, in the range $2 < p_{\rm t} < 12$ GeV/$c$. In both 
panels, the error bars (empty boxes) represent the statistical (systematic) 
uncertainties. A 3.5$\%$ normalization uncertainty is not shown. The solid 
curves are FONLL calculations and the bands display the theoretical systematic 
uncertainties.  Also shown, are the FONLL calculations and systematic 
theoretical uncertainties for muons from charm (long dashed curves) and beauty 
(dashed curves) decays. The lower panels show the corresponding 
ratios between data and FONLL calculations.}
\label{fig:xsect-int}
\end{figure}

The measured differential production cross sections of muons 
from heavy flavour decays as a function of \pt\ in the rapidity region 
$2.5 < y < 4$ and as a function of $y$ in the range 
$2 < p_{\rm t} < 12$ GeV/$c$ are displayed in Fig.~\ref{fig:xsect-int} 
(circles), left and right panels, respectively. 
The error bars (which are smaller than symbols in most of the 
\pt\ and $y$ bins) represent the statistical uncertainties. The boxes 
correspond to the systematic uncertainties. The systematic uncertainty on
$\sigma_{\rm MB}$ is not included in the boxes. 
The results are compared to FONLL predictions~\cite{FONLL1,FONLL3} 
(black curve and shaded band for the systematic uncertainty). The central 
values of FONLL calculations use CTEQ6.6~\cite{pdf} parton distribution 
functions, a charm quark mass ($m_{\rm c}$) of 1.5~GeV/$c^2$, a beauty 
quark mass ($m_{\rm b}$) of 4.75~GeV/$c^2$ and 
the renormalization ($\mu_{\rm R}$) and factorization 
($\mu_{\rm F}$) QCD scales such that 
$\mu_{\rm R}/\mu_0 = \mu_{\rm F}/\mu_0 = 1$ 
($\mu_0 = m_{\rm t,q} = \sqrt {p^2_{\rm t} + m^2_{\rm q}}$). 
The theoretical uncertainties correspond to the variation of 
charm and beauty quark masses in the ranges 
$1.3 < m_{\rm c} < 1.7$ GeV/$c^2$ and 
$4.5 < m_{\rm b} < 5.0$ GeV/$c^2$, and QCD scales in the ranges 
$0.5 <\mu_{\rm R}/\mu_0 < 2$ and $0.5 <\mu_{\rm F}/\mu_0 < 2$ with the 
constraint $0.5 <\mu_{\rm F}/\mu_{\rm R} < 2$.
The FONLL predictions for muons from beauty 
decays include the components of muons coming from direct 
b-hadron decays and from b-hadron decays via c-hadron decays 
(e.g.~B $\rightarrow$ D $\rightarrow$ $\mu$ channel). 
The uncertainty band is the envelope of the resulting cross sections. 
The ratios between data and FONLL predictions are shown 
in the bottom panels. A good description of the data is 
observed within uncertainties, for both the \pt\ distribution 
(up to 12 GeV/$c$) and the $y$ distribution (in the \pt\ range from 2 
to 12 GeV/$c$). The measured production cross sections are systematically 
larger than the central values of the model predictions. 
The ratio of data over central values of FONLL calculations 
as a function of \pt\ and $y$ is about 1.3 over the whole \pt\ and $y$ ranges. 
This is consistent with the ALICE measurements of 
the \pt -differential production cross sections  
of D mesons~\cite{Dmes11} in the central rapidity region. 
The CMS and ATLAS Collaborations made complementary measurements of the 
heavy flavour production, with electrons and/or muons measured at 
mid-rapidity in pp collisions at $\sqrt s$ = 7 TeV~\cite{cms2,atlas2}. 
The production of muons from beauty decays, measured by the CMS Collaboration 
in $\vert \eta \vert < 2.1$ and at high \pt\ ($p_{\rm t} > 6$~GeV/$c$), 
exhibits a similar agreement with NLO pQCD calculations within uncertainties: 
the data points lie in the upper limit of the model predictions. 
The results from the ATLAS Collaboration concerning the production of muons 
and electrons from heavy flavour decays in 
$\vert \eta \vert < 2.0$ (excluding $1.37 < \vert \eta \vert < 1.52$) 
and in the region $7 < p_{\rm t} < 27$~GeV/$c$ are also consistent with FONLL 
calculations.

The theoretical charm and beauty components are also displayed in 
Fig.~\ref{fig:xsect-int}. According to these predictions, 
the muon contribution from beauty decays is expected to dominate 
in the range \pt\ $\gtrsim$ 6 GeV/$c$. In this region, 
it represents about 62$\%$ of the heavy flavour decay muon cross section.

\begin{figure}[hpbt]
\centering
{\includegraphics[width=.65\textwidth]{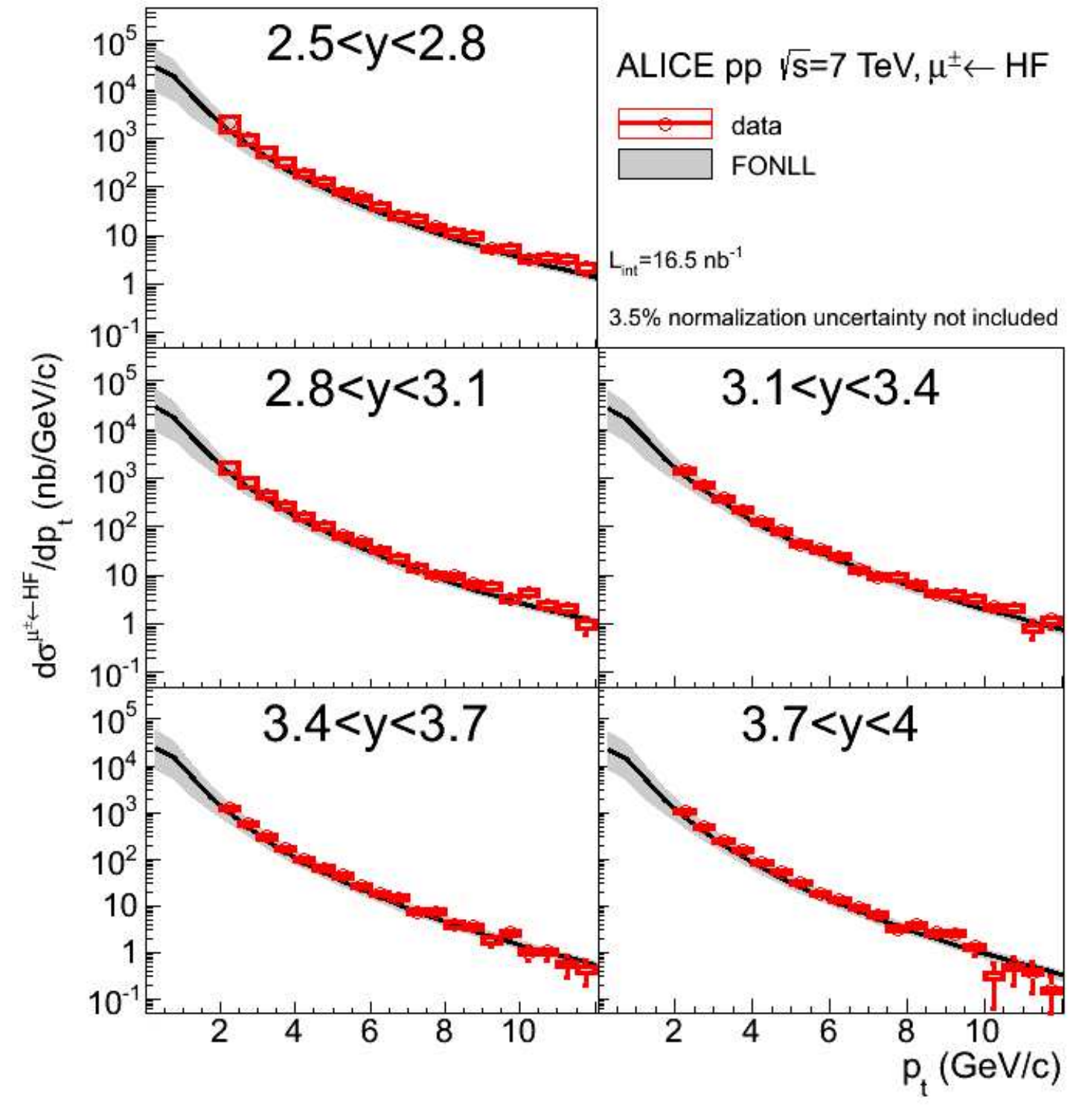}}
{\includegraphics[width=.65\textwidth]{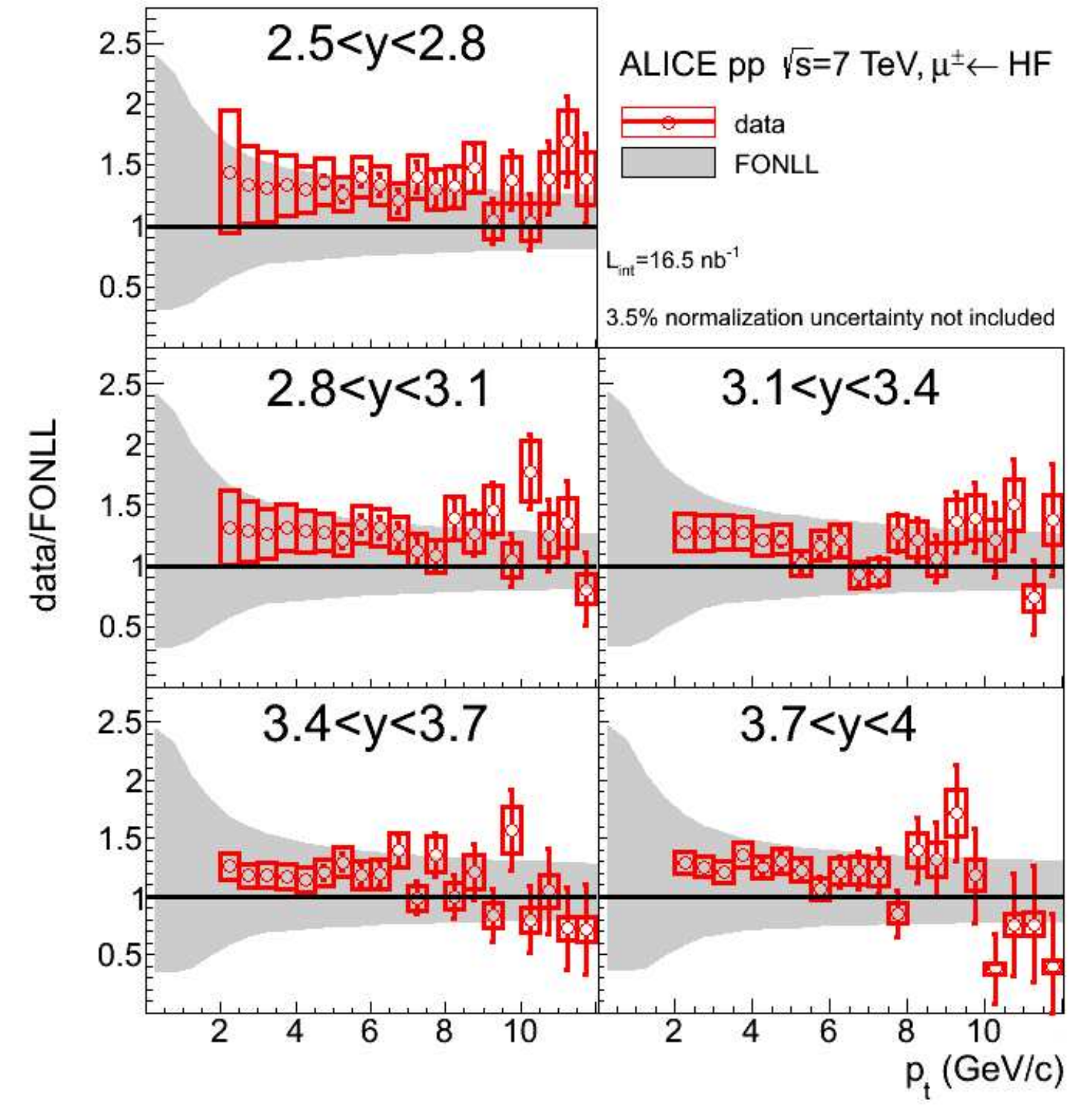}}
\caption{Upper panel: \pt -differential production cross section of muons from 
heavy flavour decays in five rapidity regions mentioned in the figures. 
The error bars (empty boxes) represent the statistical (systematic) 
uncertainties. A 3.5$\%$ normalization uncertainty is not shown. The solid 
curves are FONLL calculations and the bands display the theoretical systematic 
uncertainty. Lower panel: ratio between data and FONLL 
calculations. }
\label{fig:xsect-diff}
\end{figure}

A similar comparison between data and FONLL calculations was performed 
in five rapidity intervals from $y$ = 2.5 to $y$ = 4 
(Fig.~\ref{fig:xsect-diff}, upper panels). The corresponding ratio 
of data over FONLL predictions is depicted in the lower panels of 
Fig.~\ref{fig:xsect-diff}. The model calculations provide an overall 
good description of the data up to \pt\ = 12 GeV/$c$ in all rapidity 
intervals, within experimental and theoretical uncertainties. 

\section {Conclusions}

We have presented measurements of the differential production cross sections 
of muons from heavy flavour decays in the rapidity range $2.5 < y < 4$ 
and transverse momentum range $2 < p_{\rm t} < 12 $ GeV/$c$,
in pp collisions at $\sqrt s$ = 7 TeV with the ALICE experiment. 
The FONLL pQCD calculations are in good agreement with data within 
experimental and theoretical uncertainties, although 
the data are close to the upper limit of the model calculations. 
Both the \pt\ and $y$ dependence of the heavy flavour 
decay muon production cross section is well described by the 
model predictions. The results provide an important baseline for 
the study of heavy quark medium effects in nucleus--nucleus collisions.

%
\newenvironment{acknowledgement}{\relax}{\relax}
\begin{acknowledgement}
\section{Acknowledgements}
The ALICE collaboration would like to thank all its engineers and technicians for their invaluable contributions to the construction of the experiment and the CERN accelerator teams for the outstanding performance of the LHC complex.
\\
The ALICE collaboration acknowledges the following funding agencies for their support in building and
running the ALICE detector:
 \\
Calouste Gulbenkian Foundation from Lisbon and Swiss Fonds Kidagan, Armenia;
 \\
Conselho Nacional de Desenvolvimento Cient\'{\i}fico e Tecnol\'{o}gico (CNPq), Financiadora de Estudos e Projetos (FINEP),
Funda\c{c}\~{a}o de Amparo \`{a} Pesquisa do Estado de S\~{a}o Paulo (FAPESP);
 \\
National Natural Science Foundation of China (NSFC), the Chinese Ministry of Education (CMOE)
and the Ministry of Science and Technology of China (MSTC);
 \\
Ministry of Education and Youth of the Czech Republic;
 \\
Danish Natural Science Research Council, the Carlsberg Foundation and the Danish National Research Foundation;
 \\
The European Research Council under the European Community's Seventh Framework Programme;
 \\
Helsinki Institute of Physics and the Academy of Finland;
 \\
French CNRS-IN2P3, the `Region Pays de Loire', `Region Alsace', `Region Auvergne' and CEA, France;
 \\
German BMBF and the Helmholtz Association;
\\
General Secretariat for Research and Technology, Ministry of
Development, Greece;
\\
Hungarian OTKA and National Office for Research and Technology (NKTH);
 \\
Department of Atomic Energy and Department of Science and Technology of the Government of India;
 \\
Istituto Nazionale di Fisica Nucleare (INFN) of Italy;
 \\
MEXT Grant-in-Aid for Specially Promoted Research, Ja\-pan;
 \\
Joint Institute for Nuclear Research, Dubna;
 \\
National Research Foundation of Korea (NRF);
 \\
CONACYT, DGAPA, M\'{e}xico, ALFA-EC and the HELEN Program (High-Energy physics Latin-American--European Network);
 \\
Stichting voor Fundamenteel Onderzoek der Materie (FOM) and the Nederlandse Organisatie voor Wetenschappelijk Onderzoek (NWO), Netherlands;
 \\
Research Council of Norway (NFR);
 \\
Polish Ministry of Science and Higher Education;
 \\
National Authority for Scientific Research - NASR (Autoritatea Na\c{t}ional\u{a} pentru Cercetare \c{S}tiin\c{t}ific\u{a} - ANCS);
 \\
Federal Agency of Science of the Ministry of Education and Science of Russian Federation, International Science and
Technology Center, Russian Academy of Sciences, Russian Federal Agency of Atomic Energy, Russian Federal Agency for Science and Innovations and CERN-INTAS;
 \\
Ministry of Education of Slovakia;
 \\
Department of Science and Technology, South Africa;
 \\
CIEMAT, EELA, Ministerio de Educaci\'{o}n y Ciencia of Spain, Xunta de Galicia (Conseller\'{\i}a de Educaci\'{o}n),
CEA\-DEN, Cubaenerg\'{\i}a, Cuba, and IAEA (International Atomic Energy Agency);
 \\
Swedish Research Council (VR) and Knut $\&$ Alice Wallenberg Foundation (KAW);
 \\
Ukraine Ministry of Education and Science;
 \\
United Kingdom Science and Technology Facilities Council (STFC);
 \\
The United States Department of Energy, the United States National
Science Foundation, the State of Texas, and the State of Ohio.
\end{acknowledgement}
\newpage
%
%
\appendix
\section{The ALICE Collaboration}
\label{app:collab}

\begingroup
\small
\begin{flushleft}
B.~Abelev\Irefn{org1234}\And
J.~Adam\Irefn{org1274}\And
D.~Adamov\'{a}\Irefn{org1283}\And
A.M.~Adare\Irefn{org1260}\And
M.M.~Aggarwal\Irefn{org1157}\And
G.~Aglieri~Rinella\Irefn{org1192}\And
A.G.~Agocs\Irefn{org1143}\And
A.~Agostinelli\Irefn{org1132}\And
S.~Aguilar~Salazar\Irefn{org1247}\And
Z.~Ahammed\Irefn{org1225}
% ...
\renewcommand\labelenumi{\textsuperscript{\theenumi}~}
\section*{Affiliation notes}
\renewcommand\theenumi{\roman{enumi}}
\begin{Authlist}
\item \Adef{0}Deceased
\item \Adef{M.V.Lomonosov Moscow State University, D.V.Skobeltsyn Institute of Nuclear Physics, Moscow, Russia}Also at: M.V.Lomonosov Moscow State University, D.V.Skobeltsyn Institute of Nuclear Physics, Moscow, Russia
\item \Adef{Institute of Nuclear Sciences, Belgrade, Serbia}Also at: "Vin\v{c}a" Institute of Nuclear Sciences, Belgrade, Serbia
\end{Authlist}
\section*{Collaboration Institutes}
\renewcommand\theenumi{\arabic{enumi}~}
\begin{Authlist}
\item \Idef{org1279}Benem\'{e}rita Universidad Aut\'{o}noma de Puebla, Puebla, Mexico
\item \Idef{org1220}Bogolyubov Institute for Theoretical Physics, Kiev, Ukraine
\item \Idef{org1262}Budker Institute for Nuclear Physics, Novosibirsk, Russia
\item \Idef{org1292}California Polytechnic State University, San Luis Obispo, California, United States
\item \Idef{org14939}Centre de Calcul de l'IN2P3, Villeurbanne, France
\item \Idef{org1197}Centro de Aplicaciones Tecnol\'{o}gicas y Desarrollo Nuclear (CEADEN), Havana, Cuba
\item \Idef{org1242}Centro de Investigaciones Energ\'{e}ticas Medioambientales y Tecnol\'{o}gicas (CIEMAT), Madrid, Spain
\item \Idef{org1244}Centro de Investigaci\'{o}n y de Estudios Avanzados (CINVESTAV), Mexico City and M\'{e}rida, Mexico
\item \Idef{org1335}Centro Fermi -- Centro Studi e Ricerche e Museo Storico della Fisica ``Enrico Fermi'', Rome, Italy
\item \Idef{org17347}Chicago State University, Chicago, United States
\item \Idef{org1118}China Institute of Atomic Energy, Beijing, China
\item \Idef{org1288}Commissariat \`{a} l'Energie Atomique, IRFU, Saclay, France
\item \Idef{org1294}Departamento de F\'{\i}sica de Part\'{\i}culas and IGFAE, Universidad de Santiago de Compostela, Santiago de Compostela, Spain
\item \Idef{org1106}Department of Physics Aligarh Muslim University, Aligarh, India
\item \Idef{org1121}Department of Physics and Technology, University of Bergen, Bergen, Norway
\item \Idef{org1162}Department of Physics, Ohio State University, Columbus, Ohio, United States
\item \Idef{org1300}Department of Physics, Sejong University, Seoul, South Korea
\item \Idef{org1268}Department of Physics, University of Oslo, Oslo, Norway
\item \Idef{org1132}Dipartimento di Fisica dell'Universit\`{a} and Sezione INFN, Bologna, Italy
\item \Idef{org1315}Dipartimento di Fisica dell'Universit\`{a} and Sezione INFN, Trieste, Italy
\item \Idef{org1145}Dipartimento di Fisica dell'Universit\`{a} and Sezione INFN, Cagliari, Italy
\item \Idef{org1270}Dipartimento di Fisica dell'Universit\`{a} and Sezione INFN, Padova, Italy
\item \Idef{org1285}Dipartimento di Fisica dell'Universit\`{a} `La Sapienza' and Sezione INFN, Rome, Italy
\item \Idef{org1154}Dipartimento di Fisica e Astronomia dell'Universit\`{a} and Sezione INFN, Catania, Italy
\item \Idef{org1290}Dipartimento di Fisica `E.R.~Caianiello' dell'Universit\`{a} and Gruppo Collegato INFN, Salerno, Italy
\item \Idef{org1312}Dipartimento di Fisica Sperimentale dell'Universit\`{a} and Sezione INFN, Turin, Italy
\item \Idef{org1103}Dipartimento di Scienze e Tecnologie Avanzate dell'Universit\`{a} del Piemonte Orientale and Gruppo Collegato INFN, Alessandria, Italy
\item \Idef{org1114}Dipartimento Interateneo di Fisica `M.~Merlin' and Sezione INFN, Bari, Italy
\item \Idef{org1237}Division of Experimental High Energy Physics, University of Lund, Lund, Sweden
\item \Idef{org1192}European Organization for Nuclear Research (CERN), Geneva, Switzerland
\item \Idef{org1227}Fachhochschule K\"{o}ln, K\"{o}ln, Germany
\item \Idef{org1122}Faculty of Engineering, Bergen University College, Bergen, Norway
\item \Idef{org1136}Faculty of Mathematics, Physics and Informatics, Comenius University, Bratislava, Slovakia
\item \Idef{org1274}Faculty of Nuclear Sciences and Physical Engineering, Czech Technical University in Prague, Prague, Czech Republic
\item \Idef{org1229}Faculty of Science, P.J.~\v{S}af\'{a}rik University, Ko\v{s}ice, Slovakia
\item \Idef{org1184}Frankfurt Institute for Advanced Studies, Johann Wolfgang Goethe-Universit\"{a}t Frankfurt, Frankfurt, Germany
\item \Idef{org1215}Gangneung-Wonju National University, Gangneung, South Korea
\item \Idef{org1212}Helsinki Institute of Physics (HIP) and University of Jyv\"{a}skyl\"{a}, Jyv\"{a}skyl\"{a}, Finland
\item \Idef{org1203}Hiroshima University, Hiroshima, Japan
\item \Idef{org1329}Hua-Zhong Normal University, Wuhan, China
\item \Idef{org1254}Indian Institute of Technology, Mumbai, India
\item \Idef{org1266}Institut de Physique Nucl\'{e}aire d'Orsay (IPNO), Universit\'{e} Paris-Sud, CNRS-IN2P3, Orsay, France
\item \Idef{org1277}Institute for High Energy Physics, Protvino, Russia
\item \Idef{org1249}Institute for Nuclear Research, Academy of Sciences, Moscow, Russia
\item \Idef{org1320}Nikhef, National Institute for Subatomic Physics and Institute for Subatomic Physics of Utrecht University, Utrecht, Netherlands
\item \Idef{org1250}Institute for Theoretical and Experimental Physics, Moscow, Russia
\item \Idef{org1230}Institute of Experimental Physics, Slovak Academy of Sciences, Ko\v{s}ice, Slovakia
\item \Idef{org1127}Institute of Physics, Bhubaneswar, India
\item \Idef{org1275}Institute of Physics, Academy of Sciences of the Czech Republic, Prague, Czech Republic
\item \Idef{org1139}Institute of Space Sciences (ISS), Bucharest, Romania
\item \Idef{org27399}Institut f\"{u}r Informatik, Johann Wolfgang Goethe-Universit\"{a}t Frankfurt, Frankfurt, Germany
\item \Idef{org1185}Institut f\"{u}r Kernphysik, Johann Wolfgang Goethe-Universit\"{a}t Frankfurt, Frankfurt, Germany
\item \Idef{org1177}Institut f\"{u}r Kernphysik, Technische Universit\"{a}t Darmstadt, Darmstadt, Germany
\item \Idef{org1256}Institut f\"{u}r Kernphysik, Westf\"{a}lische Wilhelms-Universit\"{a}t M\"{u}nster, M\"{u}nster, Germany
\item \Idef{org1246}Instituto de Ciencias Nucleares, Universidad Nacional Aut\'{o}noma de M\'{e}xico, Mexico City, Mexico
\item \Idef{org1247}Instituto de F\'{\i}sica, Universidad Nacional Aut\'{o}noma de M\'{e}xico, Mexico City, Mexico
\item \Idef{org23333}Institut of Theoretical Physics, University of Wroclaw
\item \Idef{org1308}Institut Pluridisciplinaire Hubert Curien (IPHC), Universit\'{e} de Strasbourg, CNRS-IN2P3, Strasbourg, France
\item \Idef{org1182}Joint Institute for Nuclear Research (JINR), Dubna, Russia
\item \Idef{org1143}KFKI Research Institute for Particle and Nuclear Physics, Hungarian Academy of Sciences, Budapest, Hungary
\item \Idef{org18995}Kharkiv Institute of Physics and Technology (KIPT), National Academy of Sciences of Ukraine (NASU), Kharkov, Ukraine
\item \Idef{org1199}Kirchhoff-Institut f\"{u}r Physik, Ruprecht-Karls-Universit\"{a}t Heidelberg, Heidelberg, Germany
\item \Idef{org20954}Korea Institute of Science and Technology Information
\item \Idef{org1160}Laboratoire de Physique Corpusculaire (LPC), Clermont Universit\'{e}, Universit\'{e} Blaise Pascal, CNRS--IN2P3, Clermont-Ferrand, France
\item \Idef{org1194}Laboratoire de Physique Subatomique et de Cosmologie (LPSC), Universit\'{e} Joseph Fourier, CNRS-IN2P3, Institut Polytechnique de Grenoble, Grenoble, France
\item \Idef{org1187}Laboratori Nazionali di Frascati, INFN, Frascati, Italy
\item \Idef{org1232}Laboratori Nazionali di Legnaro, INFN, Legnaro, Italy
\item \Idef{org1125}Lawrence Berkeley National Laboratory, Berkeley, California, United States
\item \Idef{org1234}Lawrence Livermore National Laboratory, Livermore, California, United States
\item \Idef{org1251}Moscow Engineering Physics Institute, Moscow, Russia
\item \Idef{org1140}National Institute for Physics and Nuclear Engineering, Bucharest, Romania
\item \Idef{org1165}Niels Bohr Institute, University of Copenhagen, Copenhagen, Denmark
\item \Idef{org1109}Nikhef, National Institute for Subatomic Physics, Amsterdam, Netherlands
\item \Idef{org1283}Nuclear Physics Institute, Academy of Sciences of the Czech Republic, \v{R}e\v{z} u Prahy, Czech Republic
\item \Idef{org1264}Oak Ridge National Laboratory, Oak Ridge, Tennessee, United States
\item \Idef{org1189}Petersburg Nuclear Physics Institute, Gatchina, Russia
\item \Idef{org1170}Physics Department, Creighton University, Omaha, Nebraska, United States
\item \Idef{org1157}Physics Department, Panjab University, Chandigarh, India
\item \Idef{org1112}Physics Department, University of Athens, Athens, Greece
\item \Idef{org1152}Physics Department, University of Cape Town, iThemba LABS, Cape Town, South Africa
\item \Idef{org1209}Physics Department, University of Jammu, Jammu, India
\item \Idef{org1207}Physics Department, University of Rajasthan, Jaipur, India
\item \Idef{org1200}Physikalisches Institut, Ruprecht-Karls-Universit\"{a}t Heidelberg, Heidelberg, Germany
\item \Idef{org1325}Purdue University, West Lafayette, Indiana, United States
\item \Idef{org1281}Pusan National University, Pusan, South Korea
\item \Idef{org1176}Research Division and ExtreMe Matter Institute EMMI, GSI Helmholtzzentrum f\"ur Schwerionenforschung, Darmstadt, Germany
\item \Idef{org1334}Rudjer Bo\v{s}kovi\'{c} Institute, Zagreb, Croatia
\item \Idef{org1298}Russian Federal Nuclear Center (VNIIEF), Sarov, Russia
\item \Idef{org1252}Russian Research Centre Kurchatov Institute, Moscow, Russia
\item \Idef{org1224}Saha Institute of Nuclear Physics, Kolkata, India
\item \Idef{org1130}School of Physics and Astronomy, University of Birmingham, Birmingham, United Kingdom
\item \Idef{org1338}Secci\'{o}n F\'{\i}sica, Departamento de Ciencias, Pontificia Universidad Cat\'{o}lica del Per\'{u}, Lima, Peru
\item \Idef{org1146}Sezione INFN, Cagliari, Italy
\item \Idef{org1115}Sezione INFN, Bari, Italy
\item \Idef{org1313}Sezione INFN, Turin, Italy
\item \Idef{org1133}Sezione INFN, Bologna, Italy
\item \Idef{org1155}Sezione INFN, Catania, Italy
\item \Idef{org1316}Sezione INFN, Trieste, Italy
\item \Idef{org1286}Sezione INFN, Rome, Italy
\item \Idef{org1271}Sezione INFN, Padova, Italy
\item \Idef{org1322}Soltan Institute for Nuclear Studies, Warsaw, Poland
\item \Idef{org1258}SUBATECH, Ecole des Mines de Nantes, Universit\'{e} de Nantes, CNRS-IN2P3, Nantes, France
\item \Idef{org1304}Technical University of Split FESB, Split, Croatia
\item \Idef{org1168}The Henryk Niewodniczanski Institute of Nuclear Physics, Polish Academy of Sciences, Cracow, Poland
\item \Idef{org17361}The University of Texas at Austin, Physics Department, Austin, TX, United States
\item \Idef{org1173}Universidad Aut\'{o}noma de Sinaloa, Culiac\'{a}n, Mexico
\item \Idef{org1296}Universidade de S\~{a}o Paulo (USP), S\~{a}o Paulo, Brazil
\item \Idef{org1149}Universidade Estadual de Campinas (UNICAMP), Campinas, Brazil
\item \Idef{org1239}Universit\'{e} de Lyon, Universit\'{e} Lyon 1, CNRS/IN2P3, IPN-Lyon, Villeurbanne, France
\item \Idef{org1205}University of Houston, Houston, Texas, United States
\item \Idef{org20371}University of Technology and Austrian Academy of Sciences, Vienna, Austria
\item \Idef{org1222}University of Tennessee, Knoxville, Tennessee, United States
\item \Idef{org1310}University of Tokyo, Tokyo, Japan
\item \Idef{org1318}University of Tsukuba, Tsukuba, Japan
\item \Idef{org21360}Eberhard Karls Universit\"{a}t T\"{u}bingen, T\"{u}bingen, Germany
\item \Idef{org1225}Variable Energy Cyclotron Centre, Kolkata, India
\item \Idef{org1306}V.~Fock Institute for Physics, St. Petersburg State University, St. Petersburg, Russia
\item \Idef{org1323}Warsaw University of Technology, Warsaw, Poland
\item \Idef{org1179}Wayne State University, Detroit, Michigan, United States
\item \Idef{org1260}Yale University, New Haven, Connecticut, United States
\item \Idef{org1332}Yerevan Physics Institute, Yerevan, Armenia
\item \Idef{org15649}Yildiz Technical University, Istanbul, Turkey
\item \Idef{org1301}Yonsei University, Seoul, South Korea
\item \Idef{org1327}Zentrum f\"{u}r Technologietransfer und Telekommunikation (ZTT), Fachhochschule Worms, Worms, Germany
\end{Authlist}
\endgroup


\newpage
\begin{thebibliography}{99}
\bibitem{mnr92} M.L.~Mangano, P.~Nason, G.~Ridolfi, Nucl. Phys. 
B {\bf 373} (1992) 295.
\bibitem{FONLL1} M.~Cacciari, 
M.~Greco and P.~Nason, JHEP {\bf 05} (1998) 007.
\bibitem{d01} S.~Abachi {\it et al}. 
(D0 Collaboration), Phys. Rev. Lett. {\bf 74} 
(1995) 3548; \\
B.~Abbott {\it et al.} (D0 Collaboration), Phys. Lett. B {\bf 487} 
(2000)  264; \\
B.~Abbott {\it et al.} (D0 Collaboration), Phys. Rev. Lett. {\bf 85} 
(2000)  5068.
\bibitem{cdf1} F.~Abe {\it et al.} (CDF Collaboration), Phys. Rev. Lett. 
{\bf 71} (1993) 500; \\
F.~Abe {\it et al.} (CDF Collaboration), Phys. Rev. Lett. 
{\bf 71} (1993) 2396; \\
D. Acosta {\it et al.} (CDF Collaboration), 
Phys. Rev. D {\bf 65} (2002) 052005.
\bibitem{cdf2} D. Acosta {\it et al.} (CDF Collaboration), 
Phys. Rev. D {\bf 71} (2005) 032001; \\
A.~Abulencia {\it et al.} (CDF Collaboration), 
Phys. Rev. D {\bf 75} (2007) 012010;\\
T.~Aaltonen {\it et al.} (CDF Collaboration), 
Phys. Rev. D {\bf 79} (2009) 092003. 
\bibitem{cac02} M.~Cacciari and P.~Nason, Phys. Rev. Lett. 
{\bf 89} (2002) 122003.
\bibitem{cac04} M.~Cacciari {et al.}, JHEP {\bf 07} (2004) 033.
\bibitem{kni08} B.A. Kniehl {\it et al.}, Phys. Rev. D {\bf 77} (2008) 014011.
\bibitem{aco03} D. Acosta {\it et al.} (CDF Collaboration), Phys. Rev. Lett. 
{\bf 91} (2003) 241804.
\bibitem{cac03} M.~Cacciari and P.~Nason, JHEP {\bf 09} (2003) 006. 
\bibitem{kni06} B.A. Kniehl {\it et al.}, Phys. Rev. Lett. {\bf 96} 
(2006) 012001.
\bibitem{phe1} 
S.S.~Adler {\it et al.} (PHENIX Collaboration), 
Phys. Rev. Lett. {\bf 96} (2006) 032001; \\ 
A.~Adare {\it et al.} (PHENIX Collaboration), 
Phys. Rev. Lett. {\bf 97} (2006) 252002; \\
S.S.~Adler {\it et al.} (PHENIX Collaboration), 
Phys. Rev. D {\bf 76} (2007) 092002; \\
A.~Adare {\it et al.} (PHENIX Collaboration), 
Phys. Rev. Lett. {\bf 103} (2009) 082002.
\bibitem{sta1} B.~I.~Abelev {\it et al.} (STAR Collaboration), 
Phys. Rev. Lett. {\bf 98} (2007) 192301; \\
W.~Xie, (STAR Collaboration), PoS (DIS 2010) 182.  
\bibitem{cac05} M.~Cacciari, P.~Nason, and R.~Vogt, Phys. Rev. Lett. {\bf 95} 
(2005) 122001.
\bibitem{atlas1} J.~Kirk (ATLAS Collaboration), PoS (2010) 013.
\bibitem{lhcb1} R.~Aaij {\it et al.} (LHCb Collaboration), 
Phys. Lett. B {\bf 694} (2010) 209; \\
R.~Aaij {\it et al.} (LHCb Collaboration), 
Eur. Phys. J. C {\bf 71} (2011) 1645.
\bibitem{cms1} V.~Khachatryan {\it et al.} (CMS Collaboration), 
Phys. Rev. Lett. {\bf 106} (2011) 112001; \\
S.~Chatrchyan {\it et al.} (CMS Collaboration),
Phys. Rev. Lett. {\bf 106} (2011) 252001; \\
P.~Bellan (CMS Collaboration), arXiv:1109.2003 [hep-ex].
\bibitem{cms2} V. Khachatryan {\it et al.} (CMS Collaboration), 
JHEP {\bf 03} (2011) 090.
\bibitem{atlas2} G.~Aad {\it et al.} (ATLAS Collaboration), 
arXiv:1109.0525 [hep-ex], CERN-PH-EP-2011-108.
\bibitem{car04} F.~Carminati {\it et al.} (ALICE Collaboration), 
J. Phys. G: Nucl. Part. Phys. {\bf 30} (2004) 1517.
\bibitem{ale06} B.~Alessandro {\it et al.} (ALICE Collaboration), 
J. Phys. G: Nucl. Part. Phys. {\bf 32} (2006) 1295.
\bibitem{arm05} N.~Armesto {\it et al.}, Phys. Rev. D {\bf 71} (2005) 050427.
\bibitem{djo05} M.~Djordjevic, M.~Gyulassy and S.~Wicks, 
Phys. Rev. Lett. {\bf 94} (2005) 112301; \\
A.D.~Frawley, T.~Ullrich, R.~Vogt, Phys. Rep. {\bf 462} (2008) 125.
\bibitem{aam08} K.~Aamodt {\it et al.} (ALICE Collaboration), 
JINST {\bf 3} (2008) S08002.
\bibitem{HFmid11} A.~Dainese {\it et al.} (ALICE Collaboration), 
J. Phys. G: Nucl. Part. Phys. {\bf 38} (2011) 124032 and references therein.
\bibitem{Dmes11} B.~Abelev {\it et al.} (ALICE Collaboration), 
arXiv:1111.1553 [hep-ex].
\bibitem{FONLL3} M.~Cacciari {\it et al.}, in preparation.
\bibitem{jpsi} K.~Aamodt {\it et al.} (ALICE Collaboration), 
Phys. Lett. B {\bf 704} (2011) 442.
\bibitem{aph09} L. Aphecetche {\it et al.}, ALICE Internal Note 
ALICE-INT-2009-044, \\ 
https://edms.cern.ch/document/1054937/1. 
\bibitem{blo02} V.~Blobel and C.~Kleinwort, arXiv:hep-ex/0208021.
\bibitem{geant3} R.~Brun {\it et al.}, GEANT3 User Guide CERN, 
Data Handling Division DD/EE/841 (1985).
\bibitem{geant3b} R.~Brun {\it et al.}, CERN Program Library Long Write-up, 
W5013, GEANT Detector Description and Simulation Tool (1994). 
\bibitem{pyt1} T.~Sj\"ostrand, Comput. Phys. Commun. {\bf 82} (1994) 74.
\bibitem{pyt2} T.~Sj\"ostrand, S.~Mrenna, P.~Skands, 
JHEP {\bf 05} (2006) 026.
\bibitem{per10} P.~Z.~Skands, Phys. Rev. D {\bf 82} (2010) 074018.
\bibitem{zai07} Z.~Conesa del Valle {\it et al.} (ALICE Collaboration), Eur.~Phys.~J. C {\bf 49} (2007) 149.
\bibitem{pho} R.~Engel, J.~Ranft, S.~Roesler, 
Phys. Rev. D {\bf 52} (1995) 1459.
\bibitem{cho11} M.~Chojnacki {\it et al.} (ALICE Collaboration), 
J. Phys. G: Nucl. Part. Phys. {\bf 38} (2011) 124074.
\bibitem{gag11} M.~Gagliardi {\it et al.} (ALICE Collaboration), 
arXiv:1109.5369 [hep-ex]; \\
K.~Aamodt {\it et al.} (ALICE Collaboration), Measurement of inelastic, single 
and double diffraction cross sections in proton-proton 
collisions at LHC with ALICE, in preparation.
\bibitem{vdm} S.~van der Meer, ISR-PO/68-31, KEK68-64.
\bibitem{beam} G.~Anders {\it et al.}, CERN-ATS-Note-2011-004 PERF.
\bibitem{pdf} P.M.~Nadolsky {\it et al.}, Phys. Rev. D {\bf 78} (2008) 013004.
\end{thebibliography}
\end{document}